# Strategic Integration of AI Chatbots in Physics Teacher Preparation: A TPACK-SWOT Analysis of Pedagogical, Epistemic, and Cybersecurity Dimensions

N. Mohammadipour[1,*]

[1]Department of Physics Education, Farhangian University, P.O. Box 14665-889, Tehran, Iran

[*]Corresponding email: naser.kurd@cfu.ac.ir

## Abstract

This study investigates the strategic and epistemically responsible integration of AI-powered chatbots into physics teacher education by employing a TPACK-guided SWOT framework across three structured learning activities. Conducted within a university-level capstone course on innovative tools for physics instruction, the activities targeted key intersections of technological, pedagogical, and content knowledge (TPACK) through chatbot-assisted tasks: simplifying abstract physics concepts, constructing symbolic concept maps, and designing instructional scenarios. Drawing on participant reflections, classroom artifacts, and iterative feedback, the results highlight internal strengths such as enhanced information-seeking behavior, scaffolded pedagogical planning, and support for symbolic reasoning. At the same time, internal weaknesses emerged, including domain-specific inaccuracies, symbolic limitations (e.g., LaTeX misrendering), and risks of overreliance on AI outputs. External opportunities were found in promoting inclusive education, multilingual engagement, and expanded zones of proximal development (ZPD), while external threats included prompt injection risks, institutional access gaps, and cybersecurity vulnerabilities. By extending existing TPACK-based models with constructs such as AI literacy, prompt-crafting competence, and epistemic verification protocols, this research offers a theoretically grounded and practically actionable roadmap for embedding AI in STEM teacher preparation. The findings affirm that, when critically scaffolded, AI chatbots can support metacognitive reflection, ethical reasoning, and instructional innovation in physics education if implementation is paired with digital fluency training and institutional support.

**Keywords**: AI Chatbots; Physics Teacher Education; TPACK Framework; SWOT Analysis; AI Literacy; Cybersecurity in Education; Instructional Design; Epistemic Responsibility.

---

[1] Corresponding email: naser.kurd@cfu.ac.ir

# 1. Introduction

## 1.1 Background and Motivation

In recent years, university-level physics education has faced increasing demands for innovation in teaching methodologies, particularly in response to the growing diversity of student needs, learning styles, and technological expectations. Traditional didactic approaches often struggle to keep pace with the evolving demands of modern physics curricula, which require not only conceptual understanding but also symbolic reasoning, computational modeling, and real-world application (Redish, 2003). The integration of digital tools into the physics classroom has become a strategic imperative, particularly as students enter higher education with prior exposure to interactive and adaptive technologies in their secondary education experiences (Lai, 2022). These learners, often referred to as "digital natives," expect educational environments to mirror the interactivity and personalization they encounter in digital platforms outside academia.

Moreover, the nature of physics itself—as a discipline rooted in abstraction, mathematical formalism, and conceptual modeling (Asheghi Mehmandari, 2023)—presents unique pedagogical challenges that demand more dynamic forms of instruction. Students frequently struggle with translating physical phenomena into symbolic representations and with making meaningful connections between theory, mathematics, and experimental data (Docktor & Mestre, 2014). In response, educators and researchers have increasingly turned toward technology-enhanced learning tools to foster visualization, engagement, and active inquiry. While simulations, online laboratories, and learning management systems (Kermani et al., 2023) have become more prevalent, these tools often function as static repositories of content or rigid instructional sequences, limiting opportunities for spontaneous, dialogic learning.

Against this backdrop, artificial intelligence (AI)—particularly in the form of generative language models—has emerged as a promising frontier in educational technology. Unlike conventional tools, AI chatbots offer interactive, responsive, and adaptive communication that mirrors aspects of human tutoring. These systems have the potential to provide real-time feedback, support exploratory questioning, and promote metacognitive reflection, all of which are essential for mastering the conceptual and mathematical complexity of physics (Rastgoo et al., 2022; Zarchi et al., 2024, Moshiri et al., 2023). As such, investigating how these AI tools can be meaningfully integrated into university-level physics instruction is not only timely but also essential for preparing the next generation of educators to engage with evolving technological and epistemological landscapes.

## 1.2 The Rise of AI Chatbots in Education

One of the most transformative technological developments in recent educational discourse is the emergence of AI-powered chatbots built on large language models (LLMs). These systems—such as OpenAI's ChatGPT, Google Gemini, or Anthropic's Claude—leverage transformer-based architectures trained on massive datasets to generate human-like text, respond to natural language prompts, and simulate reasoning across a wide range of subjects (OpenAI, 2023). Unlike traditional educational software, AI chatbots offer flexible, conversational interfaces that enable learners to pose questions, explore follow-ups, clarify concepts, and receive immediate, adaptive responses. Their capacity to mimic tutorial-style dialogue has positioned them as powerful tools for both formal instruction and informal learning environments (Qadir, 2023).

In the context of education, particularly at the tertiary level, AI chatbots are being increasingly explored for applications ranging from automated writing support and reading comprehension assistance to STEM content generation and formative assessment (Zawacki-Richter et al., 2019). Their appeal lies in their ability to break the limitations of static instructional content by allowing students to interact with knowledge dynamically. Within science and engineering education, LLM-based chatbots have shown potential to explain complex processes, generate analogies, and even assist with code writing or mathematical problem-solving (Belda-Medina, & Kokošková, 2023). However, their effectiveness varies depending on domain complexity, task type, and the level of conceptual abstraction required—challenges that are particularly pronounced in fields like physics, where symbolic logic and mathematical precision are central.

Despite these challenges, AI chatbots have already made meaningful inroads into teacher education. For instance, recent research by Pernaa et al. (2023) demonstrated that AI chatbots could meaningfully support chemistry teacher candidates in information-seeking tasks, prompting critical thinking and fostering digital literacy skills. Such studies highlight the pedagogical potential of integrating conversational AI into teacher training programs—both as a tool for content support and as a medium for developing reflective, critical engagement with technology. Yet, the success of these implementations hinges on context-sensitive adaptation. Physics education, which often relies on advanced symbolic reasoning, experimental interpretation, and problem-solving, poses distinct demands that cannot be directly inferred from studies in other disciplines. This underscores the need for targeted research to investigate how AI chatbots can be responsibly and effectively utilized in physics teacher preparation.

### 1.3 Challenges in University-Level Physics Education

Physics at the university level is widely regarded as one of the most intellectually demanding disciplines, requiring students to integrate conceptual understanding, mathematical reasoning, and experimental interpretation. Despite years of pedagogical reform, significant learning obstacles persist. A large body of research has shown that students—even at advanced levels—often harbor deeply rooted misconceptions about fundamental physical principles, such as Newtonian mechanics, thermodynamics, or electromagnetism (McDermott & Redish, 1999). These misconceptions are not easily corrected through traditional lecture-based instruction, which tends to emphasize formalism over conceptual dialogue. As a result, students may learn to manipulate equations without fully grasping the physical meaning behind them, leading to superficial learning and performance gaps.

Another critical challenge is the **symbolic and mathematical complexity** inherent in university-level physics. From tensor calculus in relativity to eigenvalue problems in quantum mechanics, students are required to fluently move between natural language, diagrams, equations, and abstract representations. This multimodal cognitive demand often overwhelms learners who lack prior exposure to such integrative thinking. Furthermore, many students struggle with translating physical scenarios into mathematical formalisms and vice versa—skills essential for modeling and problem-solving (Tuminaro & Redish, 2007). Although textbooks and simulations can offer some support, they are frequently limited to static formats that do not respond adaptively to student confusion.

Additionally, there is a growing recognition that physics instruction must evolve in tandem with technological and epistemological changes in science itself. As physics becomes increasingly data-driven and interdisciplinary—intersecting with computational science, artificial intelligence, and engineering—educators are challenged to prepare students not only to understand foundational theories but also to engage critically with modern tools and digital environments (Wood, et al., 2016). However, integrating such technologies into the curriculum is complicated by infrastructure limitations, instructor preparedness, and pedagogical inertia. These factors make it difficult for departments to modernize instruction while preserving disciplinary rigor. In this context, the strategic adoption of emerging technologies—such as AI chatbots—offers a potential avenue for addressing both cognitive and institutional barriers in university-level physics education.

**1.4 Purpose of the Study**

The primary purpose of this study is to explore the strategic and responsible integration of AI-powered chatbots into university-level physics teacher education, with a specific focus on their potential to support information seeking**,** conceptual understanding**,** instructional planning**,** and epistemically safe pedagogical reasoning (Figure 1)**.** As large language models (LLMs) such as GPT-4, Claude, and Gemini become increasingly embedded in educational environments, it is essential to move beyond surface-level functionality and interrogate how these systems are actually perceived, evaluated, and utilized by future physics educators—particularly within cognitively rigorous and symbolically complex domains like physics.

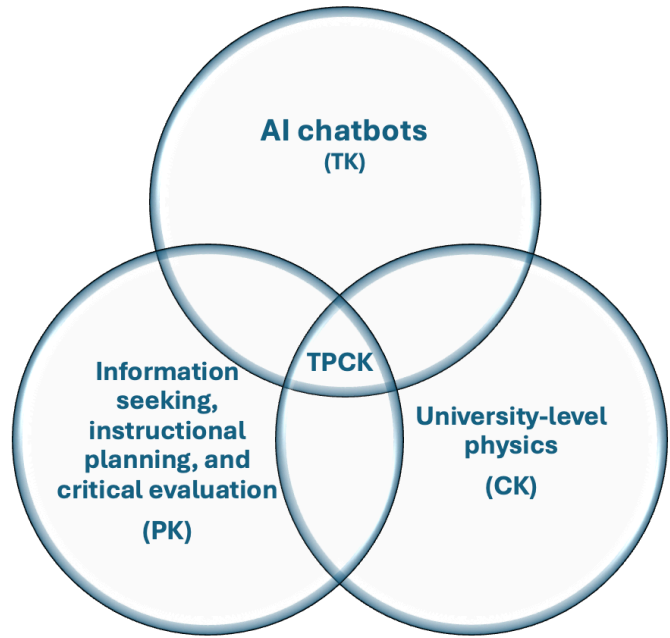

**Figure 1** illustrates the contextual framework of this study, grounded in the TPCK model. It highlights the integration of AI chatbots (technological knowledge), physics-related information seeking, instructional planning, and critical evaluation (pedagogical knowledge), and advanced university physics content (content knowledge). The intersection of these three domains represents the zone of effective teaching with AI: where digital tools are aligned with both disciplinary rigor and pedagogical purpose. This structure guides our design of learning tasks and our subsequent SWOT-based evaluation.

While previous studies have examined the use of generative AI tools in general education or in adjacent STEM fields such as chemistry (Pernaa et al., 2023), limited research has addressed the discipline-specific demands of physics education—such as vector calculus, symbolic representation (e.g., LaTeX), and the causal reasoning structures embedded in physical models. Moreover, physics teacher education uniquely intersects both content mastery **and** pedagogical competence, positioning it as an ideal context for investigating the multifaceted affordances and constraints of AI-supported instruction.

To fill this gap, the current study designed a series of structured chatbot-assisted activities that simulate authentic tasks in physics teaching—ranging from simplifying complex concepts, to constructing symbolic concept maps, to generating lesson plans using AI assistance. These activities were framed and analyzed through a TPACK-guided SWOT framework, enabling a strategic evaluation of how AI tools mediate technological (TK), pedagogical (PK), and content (CK) knowledge in pre-service physics teacher development. In adapting the TPACK model, we extend the framework introduced by Pernaa et al. (2023) in chemistry education to capture the distinctive epistemological, cognitive, and representational demands of physics instruction.

In addition to traditional TPACK elements, this study introduces AI literacy, cybersecurity resilience, and epistemic responsibility as new critical variables. These dimensions reflect the emerging need for future educators to understand how LLMs work (e.g., prompt engineering, hallucination risk), how to protect against misuse (e.g., prompt injection, adversarial behavior), and how to verify AI-generated content (e.g., triangulation with authoritative sources). By foregrounding these variables, the study contributes to a broader rethinking of what it means to prepare teachers in the age of generative AI—not just as content deliverers, but as critical evaluators, ethical practitioners, and digitally literate facilitators of learning.

The overarching research inquiry guiding the study is:

**In what ways do AI-supported information seeking, instructional planning, and critical reasoning influence the development of pedagogical practices in university-level physics teacher preparation?**

To address this question, the study employed a qualitative methodology based on thematic analysis of participant reflections, and a structured SWOT assessment to surface key internal and external factors influencing chatbot adoption. Section 4 of the paper presents findings categorized as strengths, weaknesses, opportunities, and threats, while also cross-referencing these results with conceptual constructs such as ZPD, symbolic fluency, prompt-crafting, and AI verification protocols. Section 5 then synthesizes these findings into practical and policy-relevant insights for curriculum developers, teacher educators, and institutional stakeholders.

Ultimately, this study positions AI chatbots not merely as digital tools for instruction, but as catalysts for reconfiguring knowledge construction, reasoning practices, and pedagogical ethics in physics education. It provides theoretical grounding and practical direction for embedding AI in teacher preparation programs in ways that are strategically adaptive, epistemically robust, and ethically grounded.

### 1.5 Research Questions

Grounded in the need to explore the pedagogical, ethical, and cybersecurity implications of artificial intelligence in physics education, this study is guided by the following central research question:

**RQ1:**
*How do university-level physics teacher candidates perceive and experience the use of AI chatbots in physics-related information seeking, conceptual exploration, instructional planning, and prompt-based engagement?*

To support a deeper understanding of this overarching inquiry, the study also addresses the following sub-questions:

- **RQ1.1:**
  *What strengths and weaknesses do participants identify in using AI chatbots for understanding, simplifying, and teaching advanced physics concepts—especially in terms of conceptual clarity, symbolic accuracy, and instructional alignment?*
- **RQ1.2:**
  *What opportunities and threats do participants perceive in the integration of AI chatbots into physics teacher education programs, particularly regarding inclusivity, institutional readiness, and cybersecurity vulnerabilities (e.g., prompt injection)?*
- **RQ1.3:**
  *How do participants evaluate the pedagogical usefulness, scientific reliability, and ethical and epistemic safety of chatbot-generated physics content, including issues of misinformation, overreliance, and content verification?*

These refined questions aim to capture not only the cognitive and instructional affordances of AI tools, but also their strategic, ethical, and technical implications within evolving physics education frameworks. They serve as the analytical foundation for the study's TPACK-guided SWOT evaluation and contribute to developing actionable insights for curriculum designers, teacher educators, and policymakers in AI-enhanced STEM education.

# 2. Literature review

## 2.1 Advances in AI: LLMs, Optimization, and Security Challenges

Large Language Models (LLMs) represent a transformative advancement in neural-network-driven natural language processing (NLP), capable of producing coherent, context-aware, and human-like text (Mahdavi, 2022a). Models such as OpenAI's GPT-4, Google's Gemini, and Anthropic's Claude are built using transformer architectures and are trained on massive text datasets followed by fine-tuning via supervised learning and reinforcement learning (Zarchi, & Attaran, 2017) from human feedback (e.g., PPO) to optimize for human-like responses (Hu et al., 2023). Efficient fine-tuning techniques—such as Low-Rank Adaptation (LoRA) and parameter-efficient tuning—enable domain-specific customization that enhances model precision while reducing computational overhead (Hu et al., 2023).

Alongside these breakthroughs, the integration of LLMs into real-world systems has highlighted critical cybersecurity vulnerabilities. Notable are **prompt-injection attacks**, where adversarial inputs manipulate model behavior—risking misinformation, data exposure, or unauthorized command execution (Liu et al., 2023). These vulnerabilities are increasingly relevant in educational applications when relying on conversational AI, necessitating robust defenses such as polymorphic prompts or layered input sanitization and continuous log monitoring to maintain trust and reliability (Mirnajafizadeh, 2024).

Furthermore, LLMs are now being utilized in **cybersecurity operations**, from anomaly detection to malware classification (Mirnajafizadeh, 2024), showing the potential for intelligent collaboration in threat workflows. However, adversarial behaviors—like deploying AI-powered worms or malicious prompt payloads—demonstrate the need for secure deployment protocols, especially in pedagogical environments (Sunkara, 2021).

By merging advancements in neural optimization (Mahdavi, 2022b, Ahmadi, 2023) and cybersecurity resilience (Mirnajafizadeh, 2024), this study situates AI chatbots within a framework that prioritizes not only pedagogical value but also scientific reliability and ethical safety. For physics teacher education, this means chatbot integration should be accompanied by prompt-crafting skills, content verification procedures, and safeguards against potential AI-driven misinformation or adversarial misuse.

## 2.2 Large Language Models and Educational Chatbots

Large Language Models (LLMs) represent a significant breakthrough in natural language processing (NLP), enabling machines to generate coherent, contextually relevant, and often human-like responses. These models—such as OpenAI's GPT-4, Google's Gemini, and Anthropic's Claude—are trained on massive corpora of internet data and fine-tuned using supervised learning and reinforcement learning (Mahdavi et al., 2024; Toloeia et al., 2017) from human feedback (RLHF) to simulate complex human dialogue (OpenAI, 2023). The emergence of LLM-powered chatbots has opened new frontiers in education by enabling human-AI interaction that is responsive, adaptive, and scalable.

In educational contexts, chatbots based on LLMs have been increasingly adopted as support tools for learners, offering features such as summarizing content, generating practice quizzes, simplifying difficult concepts, and simulating tutoring conversations (Zawacki-Richter et al., 2019). Their accessibility and responsiveness make them particularly valuable for informal and self-regulated learning environments, allowing students to engage with content on-demand. Empirical studies across diverse fields—including language learning, medical education, and business studies—report mixed but promising outcomes, with domain complexity and user prompting skill playing critical roles in chatbot effectiveness (Belda-Medina, & Kokošková, 2023).

In recent years, STEM education has emerged as a key domain for LLM experimentation, particularly because of the discipline's demand for logical reasoning, symbolic fluency, and structured problem-solving. Some studies have shown that LLMs can facilitate concept-level explanations in mathematics, chemistry, and engineering (Qadir, 2023; Pernaa et al., 2023). However, physics poses unique challenges due to its reliance on abstract mathematical representations (e.g., LaTeX formatting, vector calculus, differential operators) and causal modeling. While chatbots may assist in generating definitions or initial analogies, they frequently struggle with precise symbolic interpretation and accurate scientific reasoning (Gaur, & Saunshi, 2023; Birhane, 2023).

At a technical level, ongoing research has highlighted structural limitations and emerging risks associated with LLMs in educational contexts. Despite their fluency, these models are prone to hallucinations, where plausible-sounding but false or misleading content is generated. This introduces epistemic risk in high-stakes subjects like physics, where misinformation can lead to conceptual errors (Gilson et al., 2023; Van Dis et al., 2023). Moreover, prompt engineering—the practice of crafting specific inputs to optimize chatbot output—has become an essential skill for educators and students alike, but also introduces vulnerabilities. Sophisticated **prompt injection attacks** can manipulate chatbot responses to bypass guardrails, reveal confidential data, or introduce bias into learning interactions (Liu et al., 2023).

To address these challenges, researchers have proposed technical safeguards, including **polymorphic prompting** and **LoRA-based fine-tuning**, which allows educators to adapt general-purpose LLMs to specific subject domains using lightweight training methods (Hu et al., 2023). However, integrating such measures into educational practice requires not only technological expertise but also pedagogical foresight and cybersecurity literacy.

In summary, while LLM-based educational chatbots offer substantial promise for enhancing learning engagement and access, particularly in STEM fields, their effective use requires a nuanced understanding of both their capabilities and limitations. As this study shows, embedding LLMs into teacher education—particularly physics—demands deliberate scaffolding of AI literacy, critical evaluation skills, and awareness of the evolving technical landscape that shapes these tools.

## 2.3 AI in Physics Education

The integration of artificial intelligence (AI) into physics education has gained considerable traction, driven by the need to enhance students' conceptual understanding, symbolic reasoning, and instructional creativity. While early applications focused on adaptive learning platforms and automated grading systems, recent advances in generative AI—particularly large language models (LLMs)—have expanded the scope of possibilities for interactive and personalized learning (Roll & Wylie, 2016). In physics education, which demands both conceptual clarity and formal precision, AI chatbots offer unique opportunities for real-time feedback, conversational exploration, and personalized scaffolding of complex content.

One promising application of AI in physics education lies in facilitating conceptual explanations and addressing misconceptions. AI chatbots can translate abstract topics such as electromagnetic induction, quantum tunneling, or entropy into accessible language, generate analogies, and offer varied representations of the same concept. These affordances are especially valuable in early stages of learning, where understanding often hinges on the ability to explore and reframe ideas from multiple angles (Marshman, & Singh, 2015). Importantly, these tools can support learners' Zone of Proximal Development (ZPD) by providing just-in-time prompts and feedback, which can extend cognitive reach and encourage metacognitive reflection (Belda-Medina, & Kokošková, 2023).

However, significant limitations remain—particularly in tasks requiring symbolic reasoning, such as interpreting vector calculus, manipulating mathematical expressions, or rendering physics content in LaTeX. Studies have shown that LLMs often struggle with syntax-sensitive content, producing errors in operations like $\nabla \times \mathbf{B} = \mu_0 \mathbf{J}$ or misrepresenting symbolic units and differential forms (Gaur, & Saunshi, 2023; Birhane, 2023). These deficiencies can undermine learning if users uncritically adopt AI-generated solutions without proper verification. Moreover, because symbolic fluency and formal accuracy are foundational to physics expertise, reliance on AI for symbolic tasks raises concerns about superficial understanding and cognitive offloading.

In addition, the inherently **multimodal** nature of physics instruction—requiring coordination of graphs, equations, simulations, and verbal explanations—poses challenges for current AI tools, which remain largely text-based. While some AI systems can now support basic image generation or equation rendering, their capacity to synthesize and align these modalities with coherent instructional narratives remains limited (Nazaretsky et al., 2022). This gap further emphasizes the importance of embedding AI use within a framework that promotes critical thinking, cross-referencing, and pedagogical design.

Despite these constraints, AI systems have demonstrated value in supporting instructional design and resource generation for both teachers and students. Chatbots have been used to develop example problems, suggest laboratory investigations, or even co-construct lesson plans aligned with national curricula. As physics education increasingly incorporates computational elements, AI can also assist with coding simulations or analyzing sensor-based data from experiments—bridging the divide between theory and practice (Wood, et al., 2016).

Ultimately, while AI holds significant promise for enriching physics education, its use must be framed by epistemic responsibility, scientific rigor, and pedagogical intentionality. Rather than replacing disciplinary reasoning, AI tools should be positioned as assistive partners that support learners' exploration, prompt ethical reflection, and enhance access to disciplinary practices—without compromising the depth or integrity of scientific learning.

## 2.4 Pedagogical Frameworks for AI Adoption

The successful integration of artificial intelligence (AI) tools in education—especially generative AI chatbots—requires more than access to sophisticated technology; it demands a rethinking of pedagogical frameworks, teacher competencies, and ethical responsibilities. One of the most widely recognized models for guiding technology-enhanced teaching is the Technological Pedagogical Content Knowledge (TPACK) framework, which underscores the interconnectedness of content knowledge (CK), pedagogical knowledge (PK), and technological knowledge (TK) in effective instruction (Mishra & Koehler, 2006). Within this model, the incorporation of AI into physics education involves more than technical proficiency—it entails the orchestration of AI tools in ways that align with instructional objectives, cognitive development, and domain-specific epistemologies.

Generative AI systems like large language models (LLMs) introduce unique interactional dynamics: they simulate human dialogue, co-construct content, and respond to learner queries in real-time. These affordances make them potentially transformative for inquiry-based learning and formative assessment. However, their educational value hinges on AI literacy—defined as the capacity to critically engage with AI-generated content, understand model limitations, and responsibly deploy these systems in educational contexts (Belda-Medina, & Kokošková, 2023).

Pre-service teachers must be equipped not only with prompt-crafting and tool manipulation skills, but also with an understanding of how AI systems generate, approximate, or hallucinate knowledge.

This shift necessitates the development of epistemic responsibility among educators and learners. Given that AI chatbots may present inaccurate or decontextualized responses, teachers must help students evaluate AI output through triangulation, cross-referencing, and domain-specific reasoning. These practices align with broader educational goals of fostering metacognitive regulation, reflective judgment, and scientific literacy—core elements of responsible teaching in the age of generative technologies (Ertmer et al., 2012). In physics education, where epistemic trust is closely tied to formal reasoning and empirical validation, AI tools should be framed as probabilistic guides rather than authoritative sources. This view supports a more nuanced and critical engagement with technology, one that promotes inquiry rather than automation.

Equally important are the ethical and institutional dimensions of AI adoption. As AI systems become embedded in curricula, concerns emerge around bias amplification, over-reliance, academic integrity, and the automation of pedagogical judgment. Without careful design, AI tools risk replacing rather than supporting human-centered teaching. Holstein et al. (2019) argue that the successful integration of AI in classrooms depends on aligning system behavior with teachers' goals, values, and classroom realities. To that end, digital ethics—including privacy, transparency, and equitable access—must be central to any framework guiding AI deployment in teacher education programs.

In this light, pedagogical frameworks like TPACK must evolve to incorporate dimensions of AI literacy, cybersecurity awareness, and ethical reflection. Preparing future teachers for the complexities of AI-rich classrooms involves not just technical and content knowledge, but a commitment to epistemic integrity, pedagogical intentionality, and institutional responsibility. Only then can AI tools like chatbots function as meaningful partners in the educational process—amplifying rather than distorting the values of teaching and learning.

## 2.4.1 Information Seeking in Physics Education

While information seeking has been widely studied in fields such as library sciences and chemistry education, comparatively little research has focused specifically on discipline-based information-seeking practices in university-level physics. Physics, like other scientific domains, presents unique challenges for learners due to its reliance on multi-representational formats, including symbolic equations, graphical interpretations, computational simulations, experimental data, and increasingly, algorithmic code-based modeling (Figure 2). These diverse information types demand not only technical understanding but also cognitive agility—the ability to shift fluidly between conceptual, mathematical, and empirical frameworks while maintaining internal coherence. This complexity mirrors the "triplet model" in chemistry education (macro, symbolic, submicro), but in physics, it often spans verbal (explanatory), graphical (e.g., motion graphs or vector fields), symbolic (equations and derivations), and numerical (data and computation) dimensions (Tuminaro & Redish, 2007).

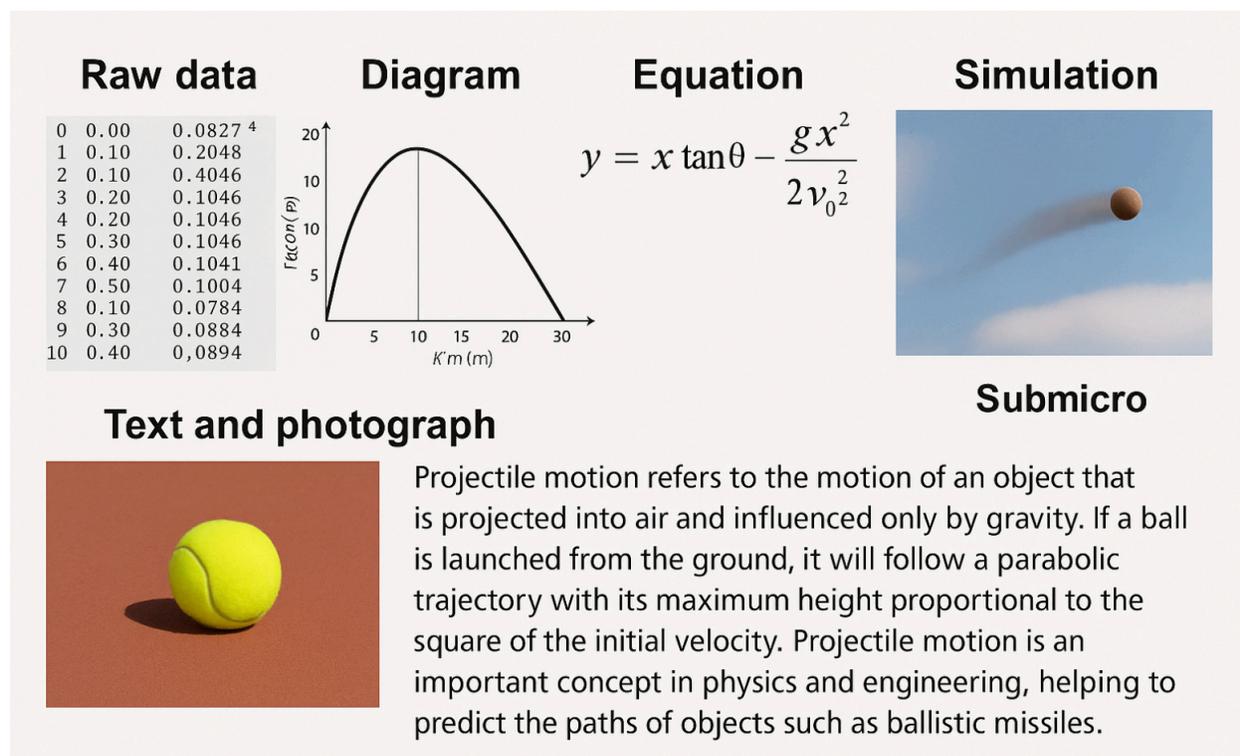

**Figure 2.** Diverse representations of projectile motion from multiple perspectives, illustrating the multifaceted nature of information in physics education.

Engaging with physics content across these modalities can result in cognitive overload, especially when students are tasked with connecting abstract theoretical constructs with empirical data or simulations. For example, understanding electromagnetic induction may require learners to synthesize a Faraday simulation, a sinusoidal current graph, and a multivariable formula—all while interpreting real-world implications. Without structured guidance, students often default to surface-level information-seeking strategies, such as keyword searching or formula memorization, rather than engaging critically with the underlying concepts. As Tuminaro and Redish (2007) observe, such learners often engage in "epistemic games"—procedural heuristics that may yield solutions but do not foster deep conceptual understanding. These issues underscore the need for explicit instruction in strategic and reflective information-seeking practices as a foundational part of physics teacher education.

The advent of AI chatbots powered by large language models (LLMs) introduces new possibilities for supporting these learning processes. Chatbots like ChatGPT offer conversational interfaces through which students can pose conceptual or procedural physics questions, receive step-by-step derivations, clarify distinctions (e.g., electric field vs. electric potential), or request help in constructing lesson materials. This interactive engagement enables learners to build personalized learning environments that are available on demand and adaptable to their zone of proximal development (ZPD), a concept rooted in Vygotskian constructivist theory. Importantly, such AI-supported inquiry can foster higher-order cognitive skills (HOCS) such as explanation, analysis, and evaluation—skills vital for future educators tasked with teaching complex scientific content.

However, AI-enhanced information seeking in physics is not without its risks. While LLMs are trained on vast datasets, they are not domain-specific experts and may produce responses with subtle but critical inaccuracies, such as misuse of physical constants, unit errors, or flawed symbolic logic. These challenges are compounded by the fact that chatbot responses often appear fluent and persuasive, making it difficult for novice users to detect errors. As a result, the integration of AI into physics education must be paired with robust training in information literacy, including prompt engineering, source verification, and cross-referencing with validated materials. Students should be encouraged to treat AI as a supportive scaffold, not as an authoritative source.

When integrated thoughtfully, AI-assisted information seeking has the potential to promote educational equity, support lifelong learning, and align with global education goals such as SDG4 (quality education for all). However, its success depends on how well teacher education programs prepare pre-service physics educators to use such tools critically and ethically. Embedding AI-based information seeking within physics education coursework—and aligning it with pedagogical frameworks such as TPACK—can empower future teachers to engage with emerging technologies in ways that enhance, rather than compromise, scientific rigor and instructional integrity.

## 2.5 Epistemic Safety, AI Literacy, and Cybersecurity in Educational AI

As generative AI systems—particularly large language models (LLMs)—become increasingly embedded in educational settings, a new set of pedagogical, epistemological, and cybersecurity responsibilities emerge. While AI tools offer powerful new ways to support inquiry, explanation, and design in teaching and learning, their adoption must be guided by critical awareness of their limitations, risks, and ethical implications. This section synthesizes emerging research across three interconnected dimensions: AI literacy, cybersecurity in generative AI, and epistemic responsibility in educational practice.

## 2.5.1 AI Literacy and Prompt-Crafting in Teacher Education

AI literacy refers not just to the technical ability to use AI tools, but to a deeper understanding of how these systems operate, what they can (and cannot) reliably do, and how to engage them responsibly within pedagogical workflows. For teacher education, this includes developing **prompt-crafting skills**—the ability to generate precise, contextually appropriate queries that elicit meaningful responses from AI chatbots.

Pre-service teachers must also understand the probabilistic nature of LLM outputs, which are generated based on statistical patterns in training data rather than true comprehension. This awareness helps mitigate overreliance and encourages users to treat chatbot outputs as suggestions rather than truths. Effective AI literacy involves knowing how to assess the appropriateness of a model for a given task, recognizing potential hallucinations or oversimplifications, and identifying when alternative tools or human expertise should take precedence. Embedding such literacy into teacher training programs ensures that AI use enhances rather than undermines instructional quality.

## 2.5.2 Cybersecurity Concerns in Generative AI

The integration of AI into education also introduces a range of cybersecurity concerns. Prompt injection attacks, where malicious input manipulates an LLM's behavior, can compromise the integrity of both content and learner interaction. Adversarial misuse, including the generation of inappropriate, biased, or misleading outputs, represents a growing challenge in AI ethics and digital safety (Mirnajafizadeh et al., 2024).

For educational contexts, these risks are especially critical: AI models may inadvertently expose learners to misinformation, reinforce stereotypes, or become vectors for academic dishonesty. Teachers must be aware of these vulnerabilities—not only to protect student data and learning outcomes, but to model safe, critical digital engagement. Institutions must also consider the infrastructural implications, including secure access, monitoring systems, and usage policies (Zarchi, & Shahgholi, 2023; Mehmandari, 2024) for AI deployment in classrooms.

Recent research has emphasized the importance of polymorphic prompt techniques and output verification layers to harden educational AI systems against manipulation and misuse (Liu et al., 2023). Incorporating these practices into teacher preparation programs contributes to broader digital resilience and safeguards academic integrity.

## 2.5.3 Epistemic Responsibility and Verification Protocols

Beyond technical literacy and cybersecurity, a central concern in AI-assisted education is epistemic responsibility—the ethical obligation to treat knowledge claims with scrutiny, especially when they are generated by non-human agents. In disciplines like physics, where accuracy, logical consistency, and empirical coherence are essential, educators and learners must develop robust habits of content verification.

These include:

- **Triangulating AI-generated content** with textbooks, academic sources, or expert opinions;
- **Identifying signs of oversimplification or symbolic inaccuracy**, particularly in mathematical or representational contexts;
- **Encouraging students to question, revise, or reject AI outputs** that fail to meet disciplinary standards.

Such verification behaviors are not only protective but pedagogically productive—they promote **critical thinking**, **scientific reasoning**, and **meta-cognitive awareness**, all of which are core goals of modern education. By training pre-service teachers to engage AI outputs with skepticism and precision, institutions can cultivate **epistemically safe classrooms**, where digital tools are used to support—not replace—judgment, inquiry, and reflection.

## 2.6 The SWOT Framework in Educational Research

The SWOT framework—which analyzes internal *Strengths* and *Weaknesses*, alongside external *Opportunities* and *Threats*—originated in strategic business planning but has since been adopted across various fields, including education, for evaluating programs, technologies, and institutional practices (Helms & Nixon, 2010). In educational research, SWOT provides a structured yet flexible lens to assess the implementation of new innovations, particularly when those innovations intersect with complex pedagogical, technological, and ethical considerations. Its utility lies in its ability to synthesize qualitative insights, stakeholder perspectives, and contextual variables into a coherent strategic evaluation—making it especially relevant in studies that explore the integration of emerging technologies like artificial intelligence.

SWOT has been increasingly applied in science education to evaluate digital learning environments, curriculum reform, and teacher development. For instance, Zawacki-Richter et al. (2019) used SWOT to analyze systemic barriers and enablers to AI adoption in higher education, identifying issues such as institutional inertia, lack of digital literacy, and potential for pedagogical innovation. More recently, Pernaa et al. (2023) employed the SWOT approach in chemistry teacher education to assess how AI chatbots support information seeking and reflection. Their findings revealed that while chatbots offered flexible access to information and supported the development of modern information literacy skills, they also presented challenges related to content accuracy, multimodal limitations, and ethical risks. The SWOT method allowed for a nuanced understanding of both technical affordances and educational implications.

In the context of this study, the SWOT framework is adopted to assess the strategic potential of AI chatbot integration in university-level physics teacher education. Unlike purely descriptive or experimental methods, SWOT enables a holistic evaluation that incorporates participant experiences, pedagogical reflections, and systemic concerns. This is particularly important in physics education, where the introduction of AI chatbots intersects with deeply held epistemic norms around precision, logic, and human reasoning. The SWOT framework thus provides an ideal structure to capture the complex interplay between innovation and discipline-specific constraints, helping inform future implementation, training, and policy development.

## 2.7 Research Gap and Contribution

While there is growing interest in the integration of artificial intelligence (AI) in higher education, existing research has largely focused on generalized educational applications such as automated writing support, language translation, and predictive analytics (Zawacki-Richter et al., 2019; Roll & Wylie, 2016). In STEM education, much of the attention has centered on adaptive platforms or data-driven tutoring systems rather than dialogic AI tools like chatbots, which engage users in real-time reasoning, iterative querying, and conversational content generation. These dialogic interactions raise unique epistemological and pedagogical questions—especially in physics education, where abstraction, precision, symbolic logic (e.g., LaTeX, vector calculus), and causal coherence are paramount.

Despite the emergence of a few subject-specific studies—such as Pernaa et al. (2023) in chemistry teacher education—there remains a clear gap in discipline-focused empirical research on AI chatbot integration within physics teacher preparation. Physics poses distinct challenges due to its reliance on mathematically rigorous representations and deep conceptual reasoning. Current language models often struggle with these representational forms, and little is known about how pre-service physics teachers critically engage with chatbot outputs, verify their epistemic validity, and incorporate them into instructional planning. Furthermore, emerging concerns around prompt-injection attacks (Mirnajafizadeh, 2024), AI misinformation, and content manipulation highlight the urgent need to address cybersecurity resilience and epistemic responsibility in teacher education contexts—concerns rarely explored in existing literature.

This study addresses these gaps by implementing a TPACK-guided SWOT analysis across three structured AI-assisted activities—focused on conceptual simplification, symbolic mapping, and instructional design. Through this framework, we not only examined pedagogical affordances and limitations of AI chatbots but also systematically assessed their scientific trustworthiness, cybersecurity vulnerabilities (Mirnajafizadeh, 2024), and ethical implications. Additionally, by embedding AI literacy, prompt-crafting skills, and verification strategies into the activity design, the study moves beyond techno-enthusiasm to critically interrogate what it means to responsibly integrate generative AI into the epistemic and pedagogical practices of future physics educators.

Thus, the study makes three key contributions:

1. **Theoretical**: It extends existing models of information seeking and TPACK by incorporating domains of AI literacy, cybersecurity resilience, and epistemic verification, offering a holistic evaluative lens for AI in education.
2. **Empirical**: It provides data-driven insights into how pre-service physics teachers interact with chatbots to construct, critique, and apply knowledge in symbolically dense and epistemologically complex domains.
3. **Practical and Policy-Oriented**: It informs curriculum design, educator training, and institutional policy by offering implementation strategies for AI in teacher education that balance innovation with pedagogical integrity and ethical safeguards.

In doing so, this study contributes to a growing body of AI-in-education research while advancing the disciplinary understanding of physics education in the age of generative AI.

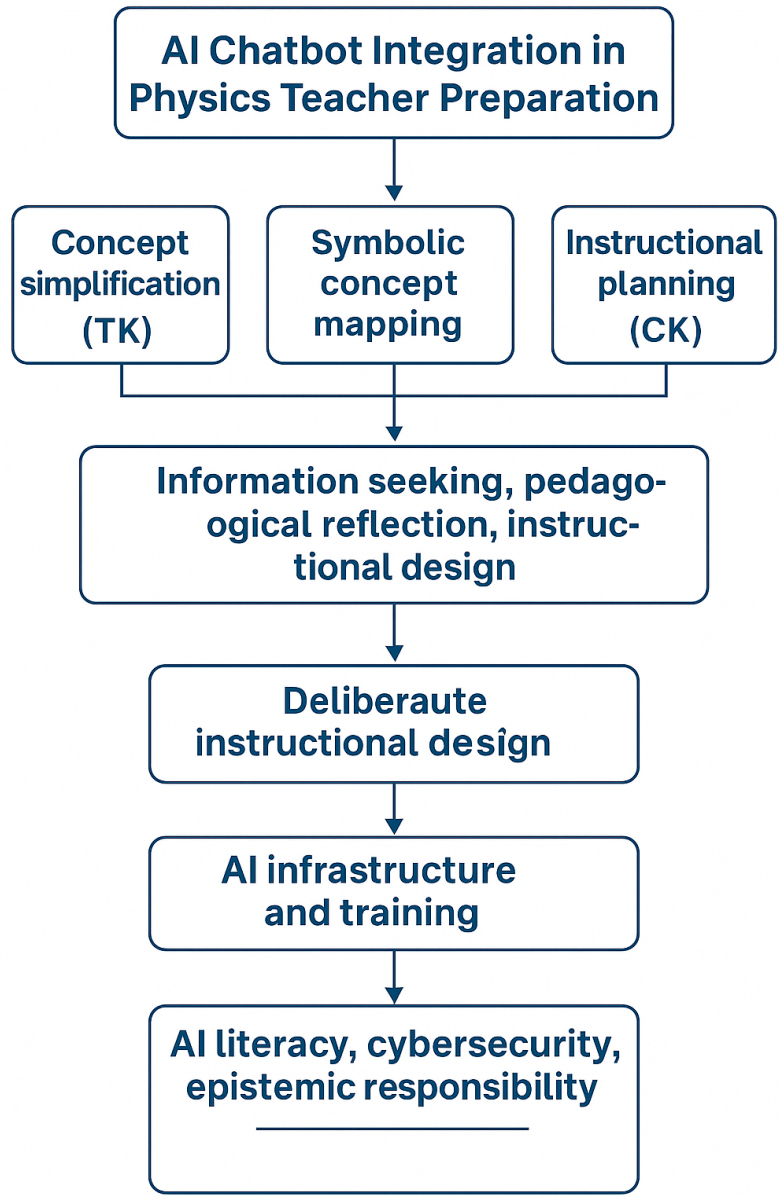

Figure 3. Research model

## 3. Methodology

### 3.1 Research Design and Theoretical Framework

This study employs a qualitative research design guided by a SWOT (Strengths, Weaknesses, Opportunities, Threats) analysis framework, enriched through the Technological Pedagogical

Content Knowledge (TPACK) model (Mishra & Koehler, 2006). The methodology was designed to explore how AI-powered chatbots can be integrated into physics teacher education, specifically through activities targeting information seeking, instructional planning, and conceptual understanding. By overlaying TPACK dimensions within each SWOT category, the study ensures both analytical depth and theoretical alignment.

### 3.1 Theoretical Framework: TPACK in Physics Education

This research is theoretically grounded in the TPACK framework (Mishra & Koehler, 2006), which provides a multidimensional structure (Zarchi, & Attaran, 2019) for understanding the intersection of technology, pedagogy, and subject content in teacher education. In the context of physics education, the framework is applied as follows:

- **Technological Knowledge (TK):** Refers to the use of AI chatbots (e.g., ChatGPT) as generative tools for instruction, planning, and knowledge construction.
- **Pedagogical Knowledge (PK):** Includes strategies for instructional planning, scaffolding, and critical evaluation, particularly in the context of information-seeking tasks.
- **Content Knowledge (CK):** Represents university-level physics, including abstract concepts such as electromagnetism, entropy, and quantum mechanics, which involve mathematical modeling and symbolic reasoning.

Intersections of these domains form hybrid knowledge areas:

- **TPK (Technological Pedagogical Knowledge):** Understanding how AI can scaffold instructional practices and information seeking.
- **TCK (Technological Content Knowledge):** Using AI to help visualize and interpret physics concepts.
- **PCK (Pedagogical Content Knowledge):** Anticipating student misconceptions and designing targeted interventions.

Despite ongoing critiques of definitional ambiguity in TPACK literature (Cox, 2008; Graham, 2011), the model's practical flexibility and integrative capacity make it an effective lens for exploring AI-based innovations in STEM education.

### 3.2 Designed AI Chatbot-Assisted Information-Seeking Activities

Three structured educational activities were designed for this study, embedded within a capstone course titled *"Innovative Tools in Physics Education"*. The course is an upper-division elective for pre-service physics teachers, emphasizing creative and critical use of emerging technologies. Each activity corresponds to distinct domains within the TPACK model and was designed to provoke reflection on cognitive and pedagogical engagement with AI tools.

### 3.2.1 Activity 1: Deconstruct a Physics Concept Using AI (PK to TPK)

In this task, students were required to select a complex physics concept—such as electromagnetic induction, entropy, or quantum tunneling—and reformulate it into an explanation suitable for a high school audience. This activity was designed to develop pedagogical translation skills while engaging critically with AI-generated explanations.

**Workflow:**

- Choose a challenging university-level physics concept.
- Use an AI chatbot (e.g., ChatGPT or Copilot) to generate a simplified explanation.
- Evaluate the explanation for clarity, scientific accuracy, and conceptual appropriateness.
- Revise the output by refining prompts and correcting errors.
- Reflect in 250–300 words on the pedagogical implications of using AI for conceptual explanation.

This activity was mapped to PK, given its focus on pedagogical transformation. However, because it incorporated conceptual evaluation and technological assistance, it also activated the TK and CK domains—thereby engaging the full TPACK framework.

### 3.2.2 Activity 2: Construct a Visual Concept Map for Symbolic Physics Content (CK to TCK)

The second activity involved creating a Novak-style concept map centered around a key topic in physics, such as Maxwell's equations or Newtonian mechanics. Emphasis was placed on the symbolic, hierarchical, and multimodal nature of physics knowledge.

**Expected features:**

- Inclusion of 15–20 key physics concepts with hierarchical structuring.
- Integration of vector diagrams, equations, LaTeX notations, and illustrative elements.
- Use of AI to support definition retrieval, conceptual linkage, and symbolic accuracy.
- Documentation of each AI contribution and critical reflection on its usefulness.

This activity primarily addressed CK but extended into TCK through the use of AI to engage with symbolic representations and their relationships. It also allowed participants to assess the chatbot's capabilities and limitations in dealing with formal physics notation.

### 3.2.3 Activity 3: Design a Physics Learning Scenario Using AI-Supported Instructional Planning (TPACK)

In the third activity, students were tasked with designing a mini instructional unit using AI tools as part of the pedagogical strategy. This task called for integrating all three TPACK knowledge domains and was intended to simulate authentic instructional design.

**Components:**

- A 2-session lesson plan (e.g., on wave interference or projectile motion), including learning goals, audience, and instructional strategy.
- A student-facing activity that incorporated chatbot use (e.g., real-time Q&A, personalized examples, or calculation checking).
- Critical discussion on the role of AI in supporting learning, formative assessment, and cognitive scaffolding.
- A preliminary SWOT analysis completed by the student based on their planning process, to be compared with the collective research-wide SWOT evaluation.

This activity constituted a fully integrated TPACK application and was the most open-ended of the three. It encouraged synthesis of knowledge, creative instructional use of AI, and critical appraisal of risks and benefits in authentic teaching scenarios.

### 3.3 SWOT Analysis Procedure

A structured SWOT framework was used to analyze the data collected from the activities above. The choice of SWOT was based on its strength in surfacing the internal affordances and external constraints of educational innovations (Helms & Nixon, 2010; Zawacki-Richter et al., 2019).

To refine this analysis and ensure theoretical depth, each quadrant was overlaid with relevant TPACK domains:

- **Strengths:** Features such as user-friendliness (TK), enhanced reflection (PK), or scaffolded content access (TPK).
- **Weaknesses:** Problems with symbolic precision, lack of context sensitivity, or oversimplification (CK, TCK).
- **Opportunities:** Emerging avenues for curriculum integration, teacher training, and personalization of learning (TPACK).
- **Threats:** Risks of epistemic confusion, over-reliance, and ethical concerns (PK, PCK).

**Data sources** included:

- Written reflections submitted by students after each activity.
- Instructor observations and field notes.
- Group discussion transcripts during debriefing sessions.

Analysis followed a two-stage process:

1. **Inductive coding** to generate themes from participant data.
2. **Deductive classification** into SWOT categories informed by theoretical alignment with TPACK domains.

### 3.4 Ethical Considerations and Research Validity

Ethical approval for the study was granted by the institutional review board of the participating university. All participants gave informed consent prior to data collection. Anonymity and confidentiality were preserved throughout the research process.

Validity was enhanced through:

- **Methodological triangulation:** Reflections, observations, and discussions were used to validate findings.
- **Collaborative coding:** Thematic agreement was ensured through peer review of codes.
- **Member checking:** Participants were invited to confirm the interpretation of their responses.

The integration of the TPACK model into SWOT analysis further reinforced both theoretical rigor and contextual specificity, enabling a rich, multidimensional understanding of AI chatbot integration in physics teacher education.

## 4. Results and Discussion

### 4.1. Deconstructing Physics Concepts with AI: Toward Pedagogical Precision

One of the central aims of Activity 1 was to assess how AI chatbots can assist pre-service physics teachers in simplifying complex university-level physics content—such as quantum tunneling or electromagnetic induction—for high school audiences. A key internal strength observed was the AI chatbot's capacity to produce clear, accessible initial explanations that students could iteratively refine. This aligns with prior research demonstrating the generative potential of large language models (LLMs) for content translation and conceptual accessibility (TPK) (OpenAI, 2023; Qadir, 2023).

From a pedagogical standpoint, students found that the chatbot's responses offered useful entry points for explaining abstract phenomena. However, they frequently identified conceptual inaccuracies, over-simplifications, or flawed analogies in the output—highlighting a persistent weakness in the model's content knowledge (CK), especially in symbolic reasoning and causality (Marshman, & Singh, 2015). These findings reflect known limitations of generative AI in handling domain-specific accuracy and formal representations (Hwang et al., 2020).

In response, students engaged in iterative prompt-crafting to clarify outputs and correct errors, fostering critical reflection and developing their pedagogical content knowledge (PCK). This represents a significant opportunity: embedding AI use into training routines not only supports reflective practice but also enhances digital literacy and epistemic agency in navigating automated tools (Belda-Medina, & Kokošková, 2023).

Moreover, students began shifting from a content-delivery mindset to one focused on metacognitive awareness—evaluating how scientific ideas are taught rather than simply transmitted. This pedagogical repositioning supports adaptive instructional design. However, the exercise also revealed external threats. Without scaffolding, students may become overly reliant on AI explanations, which risks bypassing deeper engagement with physics content. There are also broader concerns around authorship ethics, originality, and the potential for prompt-injection vulnerabilities in unsupervised AI use (Liu et al., 2023).

To mitigate these threats, physics teacher education programs should include training on responsible chatbot use, including verification procedures, ethical considerations, and cybersecurity awareness (Mirnajafizadeh et al., 2024). This ensures that chatbot integration supports both scientific integrity and instructional innovation.

SWOT Table – Activity 1: Deconstruct a Physics Concept Using AI

| Strengths | Weaknesses | Opportunities | Threats |
|---|---|---|---|
| Facilitates pedagogical translation of complex physics concepts (TPK) | Risk of inaccurate simplification or misleading analogies (CK) | Develops AI literacy and prompt-engineering competence (TPK) | Overreliance on AI-generated content may reduce deep learning (PK) |
| Enhances productivity by leveraging AI to generate simplified explanations (TK) | Requires strong prompting and iterative refinement skills (TPK) | Encourages reflective practice and iterative content improvement (PCK) | Ethical concerns around authorship, originality, and overuse (TPACK) |
| Promotes critical evaluation of scientific accuracy and clarity (PK) | May reinforce misconceptions or flawed reasoning if unverified (CK) | Promotes awareness of model limitations and trust calibration (PK) | Susceptibility to prompt-injection or adversarial misuse (TPACK, cybersecurity) |
| Builds familiarity with responsible AI integration in classroom practice (TPACK) | Lacks capacity for multimodal representation or symbolic precision (TCK) | Supports introduction of cybersecurity and verification protocols in pedagogy (TPACK) | Misalignment with standard curricula or rigid assessment frameworks (TPACK) |

## 4.2. Visualizing Symbolic Relationships through AI-Supported Concept Mapping

The second activity focused on constructing symbolic concept maps that integrated visual reasoning with domain-specific knowledge—particularly valuable in physics where abstract relationships, such as Maxwell's equations or Newtonian mechanics, require both hierarchical structuring and symbolic fluency. Pre-service teachers engaged AI chatbots to generate concise definitions, verify symbolic syntax, and suggest conceptual linkages. This reflected a TCK-centered engagement where technological tools facilitated access to content knowledge in new formats (Zawacki-Richter et al., 2019; Mishra & Koehler, 2006).

Students reported several internal strengths. The structured nature of concept mapping encouraged metacognitive reflection, enabling clearer visualization of scientific relationships (CK). Moreover, combining graphical tools with AI-supported prompts helped demystify abstract formulations—particularly among those who had previously struggled with formal mathematical representation (Hwang et al., 2020; Marshman, & Singh, 2015).

However, limitations of AI performance in symbolic accuracy emerged. Chatbots sometimes produced incorrect LaTeX formatting or misinterpreted expressions like $\nabla \times \mathbf{B} = \mu_0 \mathbf{J}$, particularly when handling vector calculus or differential equations. These symbolic errors prompted students to verify outputs by consulting textbooks or peer discussions—reinforcing triangulated information behavior, an important trait of responsible AI use (Belda-Medina, & Kokošková, 2023; Gilson et al., 2023).

This need for verification also exposed a cybersecurity and reliability layer often overlooked in pedagogical design. Chatbots can be vulnerable to adversarial prompts or misinformation when deployed without safeguards, especially when handling symbolic data that may include embedded payloads or flawed logic (Liu et al., 2023). Embedding defenses such as polymorphic prompt techniques or layered input filters into educational environments is essential (Mirnajafizadeh et al., 2024).

Still, the activity surfaced valuable opportunities. Students developed cross-verification skills and deeper engagement with physics content by actively navigating AI limitations. For those apprehensive about symbolic formalism, AI-assisted scaffolding reduced cognitive barriers—transforming what might have been a demotivating experience into one of curiosity and resilience.

Nonetheless, several external threats remain. These include unequal access to reliable visualization tools, inconsistent instructor preparedness, and cognitive overload due to symbolic density. To support ethical and effective implementation, physics education programs should ensure scaffolded instruction, secure chatbot deployment, and integrated verification procedures within teacher education curricula.

SWOT Table – Activity 2: Construct a Visual Concept Map for Symbolic Physics Content

| Strengths | Weaknesses | Opportunities | Threats |
|---|---|---|---|
| **Encourages structured thinking and conceptual clarity (CK)** | AI may struggle with accurate symbolic | Enhances multimodal and accessible learning | Cognitive overload due to complexity of symbolic content (CK) |

| | formatting and advanced notation (TCK) | in abstract physics domains (TCK) | |
|---|---|---|---|
| **Combines visual and symbolic reasoning with AI assistance (TCK)** | Overdependence on AI output without validation may introduce errors (CK) | Promotes triangulated verification and responsible AI use (TPK) | Disparities in student access to visualization tools or secure AI platforms (TPACK) |
| **Supports construction of hierarchical and relational knowledge maps (CK)** | Challenges in reviewing and interpreting AI-generated visual content (TCK) | Reduces apprehension toward abstract formalism through AI-assisted scaffolding (TPACK) | Vulnerability to prompt-injection attacks or symbolic misinformation |
| **Stimulates metacognitive engagement through symbolic reasoning (PCK)** | May fail to detect symbolic inconsistencies in adversarial prompts (TPACK, cybersecurity) | Supports integration of cybersecurity and ethical reasoning into symbolic learning activities (TPACK) | Instructor training gaps in AI-supported symbolic representation (TPACK) |

### 4.3. Designing AI-Enhanced Instructional Scenarios: Fostering TPACK Synergy

The third activity challenged participants to design a full instructional scenario incorporating AI chatbots into the physics teaching workflow. This project-based design task activated the entire TPACK framework—requiring integration of content knowledge (CK), pedagogical strategy (PK), and technological tools (TK). By asking students to simulate classroom planning, this activity brought AI use into an authentic instructional context, thereby emphasizing practical synergy among the knowledge domains (Mishra & Koehler, 2006; Graham, 2011).

Students found that AI chatbots offered strong support in generating lesson structures, producing formative questions, and proposing student-centered activities. A notable internal strength was the chatbot's capacity to assist in basic coding tasks, such as debugging Python code for kinematic simulations or assisting with Arduino-based experimental design—extending technological engagement beyond text generation into hardware-relevant planning (TK, TCK). These findings echo emerging literature showing AI's role in facilitating both instructional and technical creativity in STEM domains (Belda-Medina, & Kokošková, 2023; Cooper, 2023).

Despite these benefits, students frequently reported that AI outputs lacked essential pedagogical components such as scaffolding, formative assessment cues, and anticipatory strategies for student misconceptions. These omissions revealed a limitation in AI's alignment with pedagogical content knowledge (PCK), highlighting the need for teacher oversight and critical refinement of AI-generated lesson materials (Lai, 2022; Docktor & Mestre, 2014).

Opportunities emerged in the form of increased instructional autonomy and design-based metacognition. Several participants reported that chatbots helped expand their zone of proximal development (ZPD) by offering immediate feedback or novel instructional formats—facilitating exploratory learning and promoting just-in-time support for lesson innovation (Vygotsky, 1978; Roll & Wylie, 2016).

However, the integration of LLMs into instructional design also raised important cybersecurity and ethical considerations. Without proper safeguards, AI chatbots used in lesson planning could be vulnerable to prompt-injection attacks or generate misleading pedagogical content (Liu et al., 2023). Students may unknowingly introduce adversarial prompts that affect lesson quality, or rely on output that embeds epistemic bias or factual inaccuracies. These threats necessitate the inclusion of AI safety training in teacher education—especially prompt-crafting, content verification, and ethical analysis procedures (Mirnajafizadeh et al., 2024).

External threats also include unequal access to secure and reliable AI tools, lack of instructor readiness, and institutional hesitancy to embed generative AI into curriculum planning. Without systemic support and digital equity frameworks, the transformative potential of AI in instructional design may be limited to isolated innovations rather than mainstream adoption.

SWOT Table – Activity 3: Design a Physics Learning Scenario Using AI-Supported Instructional Planning

| Strengths | Weaknesses | Opportunities | Threats |
|---|---|---|---|
| Engages full TPACK knowledge domains in authentic instructional design (TPACK) | High cognitive demand due to complexity of design tasks (TPACK) | Fosters innovation in teaching practices using emerging technology (TPACK) | Inconsistent institutional guidance on AI use in education (TK) |
| Supports creativity and autonomy in lesson planning (PK, TK) | Limited pedagogical alignment of chatbot-generated content (PCK) | Provides a basis for personalized and differentiated learning scenarios (PCK) | Variability in student access, digital literacy, and platform reliability (TPACK) |
| Facilitates technical innovation through code support and simulation planning (TK) | Vulnerable to AI misinformation or prompt-injection risks without safeguards (cybersecurity, TPK) | Encourages development of prompt-crafting and verification skills (TPK, cybersecurity) | Risk of de-skilling in traditional instructional design processes (PK) |
| Promotes reflection on AI ethics and human-AI collaboration in classroom contexts (TPK) | Inexperience with AI tools may hinder lesson quality and coherence (TPACK) | Supports long-term AI integration strategies through pedagogical planning (TPACK) | Ethical concerns regarding authorship, transparency, and AI autonomy in lesson creation (TPK) |

## 4.4 Synthesis of AI Chatbot Use in Physics Teacher Preparation: TPACK-Based SWOT Summary

The synthesized SWOT table (Table 4) offers a consolidated analysis of the pedagogical and strategic insights derived from all three AI-assisted activities. Framed within the TPACK model, this synthesis captures the dynamic interplay between content, pedagogy, technology, and now—digital ethics and cybersecurity. As large language models (LLMs) are increasingly embedded in educational ecosystems, awareness of both pedagogical benefits and security risks becomes essential (Hu et al., 2023; Liu et al., 2023).

### 4.4.1 Internal Possibilities

Among the internal strengths, AI chatbots consistently supported the simplification of complex physics phenomena, providing initial scaffolding that participants could refine into student-appropriate explanations (TPACK) (Brandtzaeg & Følstad, 2017; Cooper, 2023). Participants used chatbots to explore diverse information-seeking paths—testing, refining, and querying concepts—which deepened technological-pedagogical engagement (TPK) (Zawacki-Richter et al., 2019). Symbolic reasoning was also enhanced as chatbots helped interpret equations and link formal concepts with instructional strategies (TCK) (Gugagayanan, 2022), while planning activities encouraged critical reflection and design-based learning (PCK) (Pernaa et al., 2023).

Furthermore, students benefited from AI-assisted simulation planning and code generation—skills aligned with neural optimization techniques like parameter-efficient tuning (Hu et al., 2023; Mahdavi, 2022b). These features supported fast iteration and exploration in both conceptual and technical dimensions of teaching.

### 4.4.2 Internal Challenges

Yet internal weaknesses persisted. AI-generated content frequently contained factual inaccuracies, misapplied analogies, or syntactic errors in physics notation (CK, TCK) (Gilson et al., 2023; Van Dis et al., 2023). Many participants noted the chatbot's limitations in multistep reasoning or in creating accurate diagrams—hindering its ability to support visualization-heavy tasks (Hwang et al., 2020). These challenges demanded significant subject-matter expertise to identify and correct AI outputs, raising concerns about cognitive overload and equity (Cox, 2008).

More critically, the possibility of adversarial prompt-injection attacks during content generation—where chatbots could be manipulated to produce misleading or ethically problematic outputs—was identified as a latent risk (Liu et al., 2023). Though not directly experienced in this study, the need for secure prompt design, input sanitization, and AI safety protocols was underscored, especially for open-ended lesson planning and real-time student interaction.

### 4.4.3 External Opportunities

From an external perspective, chatbots offered promising avenues for inclusive and multilingual education through automatic translation, simplified rephrasing, and content personalization (TPK, TPACK) (Pedro et al., 2019). Such tools can empower diverse learners and support teacher training models that embed digital competence and AI literacy (Zawacki-Richter et al., 2019). The study also revealed that chatbots promoted reflection on source trustworthiness and intellectual ownership—developing ethical awareness and professional identity in future educators (PK) (Ertmer et al., 2012).

These affordances align with global educational trends that call for both digital fluency and adaptive pedagogical capacity in AI-integrated contexts (Helms & Nixon, 2010;).

### 4.4.4 External Threats

However, the effective deployment of these tools hinges on systemic factors. Participants highlighted the need for institutional backing—including secure infrastructure, reliable access to AI platforms, and formal training in AI usage and verification (TK, TPACK). In the absence of such support, inconsistent access and uneven digital readiness may exacerbate educational inequalities.

Additionally, unchecked reliance on AI may reduce independent instructional planning and critical physics reasoning—risking de-skilling and promoting shallow engagement (PCK) (Graham, 2011). Long-term threats also include software costs, evolving licensing models, and ethical concerns related to student data privacy and chatbot autonomy (Van Dis et al., 2023; Sunkara, 2021).

Table 4. Summary of the synthesized possibilities and challenges categorized via SWOT and reflected in the TPACK framework

|  | Possibilities | Challenges |
|---|---|---|
| Internal (Within the course or activity design) | **Strengths**<br><br>– Helps explain complex physics ideas using AI scaffolds (TPACK)<br><br>– Encourages diverse information-seeking and prompt engineering (TPK)<br><br>– Supports symbolic reasoning with equations and formulas (TCK)<br><br>– Aids lesson planning, coding, and critical reflection (PCK, TK) | **Weaknesses**<br><br>– AI outputs may contain physics inaccuracies or oversimplifications (CK)<br>– Requires prior content knowledge to verify output quality (CK)<br><br>– Struggles with diagrammatic and multilevel reasoning (TCK)<br><br>– Risk of adversarial prompts or unintended AI behavior (TPK, cybersecurity) |
| External (Broader teaching environment) | **Opportunities**<br><br>– Enables inclusive learning through AI translation and rephrasing (TPK)<br><br>– Aligns with digital literacy in teacher education reforms (TPACK)<br><br>– Encourages critical thinking and ethical AI evaluation (PK)<br><br>– Potential for curriculum-level AI integration (TPACK) | **Threats**<br><br>– Needs infrastructure, AI literacy training, and ethical guidance (TK)<br><br>– Uneven digital access or institutional support may limit scalability (TPACK)<br><br>– Overuse may reduce student initiative and critical thinking (PK, PCK) |

| | | – AI tools may introduce long-term sustainability and safety issues (TPACK, cybersecurity) |
|---|---|---|

**Strategic Implications**

This synthesis affirms that AI chatbot integration into physics teacher education is both promising and complex. The updated SWOT-TPACK model highlights the importance of balancing AI's instructional potential with pedagogical integrity, content verification, and cybersecurity resilience. Embedding prompt-crafting skills, ethical reflection, and institutional safeguards into teacher preparation programs is essential for cultivating adaptive and responsible educators in the AI era. The interplay of these internal and external factors suggests that while AI chatbot integration holds transformative potential, its success depends on deliberate instructional design, sustained institutional support, and a strong emphasis on digital literacy. The TPACK-guided SWOT framework enabled us to locate these findings precisely within intersecting domains of teacher knowledge, ensuring that future curricular innovations can be both theoretically grounded and pedagogically actionable.

## 5. Discussion

The TPACK-guided SWOT analysis conducted in this study revealed both promising opportunities and significant challenges in integrating AI chatbots into physics teacher preparation programs. By designing and implementing three chatbot-assisted instructional activities—each targeting a distinct TPACK configuration—we were able to evaluate how AI tools mediate information-seeking, instructional planning, and critical pedagogical reflection in a physics education context.

At the internal level, AI chatbots demonstrated notable strengths, particularly in simplifying complex physics topics, supporting conceptual mapping, and expediting the design of physics lessons and tasks (TPK, PCK). These tools enabled student teachers to engage in generative tasks such as translating entropy or electromagnetic induction into high school-appropriate explanations, visualizing symbolic relationships using concept maps, and drafting lesson plans enriched with interactive AI-supported scaffolds. These findings align with recent research noting the productive role of AI in supporting information behavior and pedagogical reflection in educational settings (Brandtzaeg & Følstad, 2017; Zawacki-Richter et al., 2019; Nazaretsky et al., 2022).

However, internal weaknesses were also evident. Students frequently encountered conceptual inaccuracies, especially in AI-generated equations or symbolic representations—highlighting the limitations of large language models in interpreting domain-specific syntax like LaTeX or vector calculus (Gaur, & Saunshi, 2023; Birhane, 2023). This necessitated verification through triangulation with textbooks or expert consultation, which, while educationally beneficial, placed additional cognitive demands on learners (Savolainen, 2005). These findings further underscore the importance of embedding **prompt-crafting skills**, **content verification procedures**, and **AI-critical literacy** into physics teacher training.

Externally, chatbot integration revealed considerable opportunities for enhancing inclusivity and access in teacher education. Translation capabilities facilitated engagement with non-English physics texts, and the conversational interface supported differentiated learning and the expansion of students' zone of proximal development (ZPD) (Vygotsky, 1978; Belda-Medina, & Kokošková, 2023). These affordances also align with Sustainable Development Goal 4 (SDG 4), which emphasizes inclusive and equitable quality education (UNESCO, 2021). Moreover, the dynamic use of chatbots encouraged students to reflect on their instructional choices and to plan lessons with real-time feedback loops—highlighting the role of generative AI as both a cognitive tool and a reflective partner.

Nonetheless, several external threats remain. Overreliance on AI tools could diminish students' independent reasoning and symbolic fluency if not properly scaffolded—a concern echoed in broader critiques of digital outsourcing in education (Ertmer et al., 1999; Graham, 2011). Institutional constraints, such as lack of access to reliable AI infrastructure or limited instructor expertise in prompt engineering, also risk impeding successful implementation. Additionally, as recent research on **prompt-injection attacks** and adversarial misuse has shown, educational deployment of AI tools must be accompanied by robust cybersecurity measures to guard against misinformation, data leakage, or manipulation (Liu et al., 2023).

To address these challenges, we recommend that physics teacher education programs adopt a **multi-pronged strategy**:

- **Incorporate AI literacy** as a core component of the curriculum, including modules on prompt design, content verification, and ethical AI use;
- **Embed cybersecurity resilience**, including awareness of prompt-injection vulnerabilities and safe deployment protocols;
- **Promote critical AI reflection**, where learners are trained to evaluate chatbot outputs across scientific, pedagogical, and ethical dimensions;
- **Ensure institutional support**, including access to sustainable and open-source AI platforms and professional development for instructors.

Ultimately, this study positions AI chatbot integration not as a mere technological enhancement but as a catalyst for rethinking instructional design, disciplinary engagement, and epistemic responsibility in science education. Through the lens of the updated TPACK-SWOT model, we argue that preparing reflective, adaptive, and AI-literate educators is essential to harness the transformative potential of these tools in a rapidly evolving digital landscape.

## 6. Conclusion

This study explored how AI chatbots can be strategically and responsibly integrated into physics teacher preparation using a TPACK-guided SWOT framework. Through the design and analysis of three chatbot-assisted activities—focused on concept simplification, symbolic concept mapping, and instructional planning—we examined how pre-service physics teachers engage with AI tools across information seeking, pedagogical reflection, and instructional design.

The findings confirm that all three research questions were comprehensively addressed. Participants identified distinct **strengths and weaknesses** (RQ1.1), such as AI's ability to simplify complex content, scaffold pedagogical thinking, and support visual-symbolic reasoning—alongside limitations in symbolic accuracy, scientific precision, and pedagogical misalignment. **Opportunities and threats** (RQ1.2) emerged around inclusivity, multilingual access, and AI-enabled instructional innovation, but were counterbalanced by concerns over institutional support, unequal access, and cybersecurity vulnerabilities—particularly related to **prompt injection and model manipulation risks**. Participants also critically evaluated the **pedagogical value, scientific trustworthiness, and ethical implications** of AI-generated content (RQ1.3), emphasizing the importance of **prompt-crafting skills**, **epistemic caution**, and **content verification protocols** to prevent misinformation or overreliance.

Overall, the results suggest that when critically and reflectively employed, AI chatbots can enhance higher-order thinking, conceptual clarity, and instructional planning among pre-service physics teachers. However, **technological integration alone is insufficient**. Meaningful implementation requires: 1) Deliberate instructional design, 2) Institutional investment in AI infrastructure and educator training, 3) And a curriculum-level commitment to **AI literacy, cybersecurity resilience, and epistemic responsibility**.

By extending traditional models of information-seeking to include **instructional, ethical, and verification-based dimensions**, this study offers both theoretical grounding and actionable guidance for embedding AI in teacher education. Preparing future educators to engage with AI tools is not only a **technological imperative**, but a **pedagogical and ethical necessity**—crucial for cultivating **adaptive, reflective, and digitally responsible educators** in the age of generative AI.

## 7. Future Directions

To build on the findings of this study, future research should investigate the integration of AI chatbots across more diverse educational settings, cultural contexts, and larger cohorts of pre-service teachers. Mixed-methods or longitudinal designs incorporating **quantitative measures of AI literacy**, **instructional efficacy**, **self-efficacy**, and **critical reasoning** will be valuable in validating and scaling these results.

In addition, emerging challenges—such as **prompt-injection attacks**, **algorithmic bias**, and **epistemic unreliability**—highlight the urgent need for **interdisciplinary research** at the intersection of **AI, cybersecurity, and science education**. Studies should explore how these technological and ethical dimensions affect teacher cognition, instructional planning, and the formation of professional identity in STEM disciplines.

Finally, future work should address curriculum-level interventions that embed **AI ethics**, **prompt-crafting skills**, and **digital trust protocols** into teacher preparation programs. Collaborations between educators, computer scientists, and instructional designers will be essential to developing sustainable, secure, and pedagogically robust frameworks for AI-enhanced science education.